%% file: Letter.tex
\documentstyle[aps,prl,epsfig,multicol]{revtex} 
\input{defTG}

\begin{document}

\title{Entanglement between photons and atoms coupled out from a Bose-Einstein-Condensate}
\author{T.~Gasenzer, D.C.~Roberts, and K.~Burnett}
\address{Clarendon Laboratory, Department of Physics, University of Oxford,
Oxford OX1~3PU, United Kingdom\\(\today)}
\maketitle

\begin{abstract} 
We study the limitations to the relative number squeezing between photons and atoms coupled out from a homogeneous Bose-Einstein-Condensate. 
We consider the coupling between the translational atomic states by two photon Bragg processes, with one of the photon modes involved in the Bragg process in a coherent state, and the other initially unpopulated. We start with an interacting Bose-condensate at zero temperature and compute the time evolution for the system. We study the squeezing, i.e.\ the variance of the occupation number difference between the second photon and the atomic c.m.~mode. We discuss how collisions between the atoms and photon rescattering affect the degree of squeezing which may be reached in such experiments.
\\[3pt]
\end{abstract}
\begin{multicols}{2}

\input{Textbody}

This work has been supported by the A.v. Humboldt-Foundation and a Marie Curie Fellowship of the European Community under contract no.~HPMF-CT-1999-0023 (T.G.), the Marshall Trust (D.C.R.), and the United Kingdom Engineering and Physical Sciences Research Council. T.G.~would like to thank J.~Ruostekoski for useful discussions and suggestions.

\end{multicols}
\end{document}

%% file: defTG.tex
\def\vecgr#1{\mathchoice{\mbox{\boldmath$\mathrm\displaystyle#1$}}
{\mbox{\boldmath$\mathrm\textstyle#1$}}
{\mbox{\boldmath$\mathrm\scriptstyle#1$}}
{\mbox{\boldmath$\mathrm\scriptscriptstyle#1$}}}

\def\vec#1{\mathchoice{\mathrm{\mathbf{\displaystyle#1}}}
{\mathrm{\mathbf{\textstyle#1}}}
{\mathrm{\mathbf{\scriptstyle#1}}}
{\mathrm{\mathbf{\scriptscriptstyle#1}}}}
\newcommand{\bi}{\begin{itemize}}
\newcommand{\ei}{\end{itemize}}

\newcommand{\be}{\begin{equation}}
\newcommand{\ee}{\end{equation}}
\newcommand{\ba}{\begin{array}}
\newcommand{\ea}{\end{array}}
\newcommand{\bea}{\begin{eqnarray}}
\newcommand{\eea}{\end{eqnarray}}

\newcommand{\bc}{\begin{center}}
\newcommand{\ec}{\end{center}}
\newcommand{\bslide}{\begin{slide}}
\newcommand{\eslide}{\end{slide}}

\newsavebox{\TRS}
\sbox{\TRS}{\hspace{.5em} = \hspace{-1.8em}
                 \raisebox{1ex}{\mbox{\scriptsize TRS}} }

\newsavebox{\defgleich}
\sbox{\defgleich}{\ :=\ }

\newsavebox{\LSIM}
\sbox{\LSIM}{\raisebox{-1ex}{$\ \stackrel{\textstyle<}{\sim}\ $}}
\newcommand{\lsim}{\usebox{\LSIM}}
\newsavebox{\GSIM}
\sbox{\GSIM}{\raisebox{-1ex}{$\ \stackrel{\textstyle>}{\sim}\ $}}
\newcommand{\gsim}{\usebox{\GSIM}}

\newcommand{\lk}{\left}
\newcommand{\rk}{\right}
\newcounter{saveeqn}

\newcommand{\sssty}{\scriptscriptstyle}
\newcommand{\ra}{\,\rangle}
\newcommand{\la}{\langle\,}

\newcommand{\sech}{\mbox{sech}\,}

\newcommand{\Tr}{\mbox{Tr$\,$}}

\newcommand{\eh}{\mbox{$\frac{1}{2}$}}
\newcommand{\edzi}{\mbox{$\frac{1}{2i}$}}
\newcommand{\ev}{\mbox{$\frac{1}{4}$}}

\newcommand{\alp}{\alpha}

\newcommand{\del}{\delta}
\newcommand{\Del}{\Delta}

\newcommand{\lam}{\lambda}

\newcommand{\ome}{\omega}
\newcommand{\Ome}{\Omega}

\newcommand{\oa}{\hat{a}}

\newcommand{\oadg}{\hat{a}^\dagger}

\newcommand{\obdg}{\hat{b}^\dagger}
\newcommand{\ob}{\hat{b}}

\newcommand{\obet}{\hat{\beta}}

\newcommand{\obetdg}{\hat{\beta}^\dagger}

\newcommand{\oH}{\hat{H}}

\newcommand{\oHaf}{\oH_{\sssty\mathrm af}}

\newcommand{\oJ}{\hat{J}}
\newcommand{\oJv}{\hat{\Jv}}

\newcommand{\on}{\hat{n}}

\newcommand{\oS}{\hat{S}}

\newcommand{\cL}{{\cal L}}
\newcommand{\cM}{{\cal M}}

\newcommand{\ometil}{\tilde\ome}
\newcommand{\Ometil}{\widetilde\Ome}

\newcommand{\dv}{\vec{d}}

\newcommand{\cEv}{\vecgr{\cal E}}

\newcommand{\Jv}{\vec{J}}

\newcommand{\pv}{\vec{p}}

\newcommand{\rv}{\vec{r}}

\newcommand{\iKvar}[1]{{\sssty (#1)}}

%% file: Textbody.tex
%
%
%
%
%
\setcounter{equation}{0}
Producing entanglement with macroscopic samples of atoms has become an increasingly important goal for experimental studies with Bose-Einstein-Condensates (BECs). Its achievement would, in particular, be of immense value in quantum information processing (for reviews cf.~\cite{QIT}), for Heisenberg limited clocks \cite{Huelga97} and for tests of quantum state reduction \cite{Penrose98}. Techniques for the creation of entangled photon pairs \cite{EntPhotons} are already available for quantum cryptography  and teleportation \cite{QCryptQTelep}, analogous schemes involving macroscopic samples of atoms are still being sought. There are good reasons for doing this: whereas photons may be transmitted easily over large distances, atoms have the advantage of being more easily stored. 

In \cite{Roberts01} we have investigated the possibilities for exciting squeezed atoms from condensates using light scattering. 
In this letter we study the entanglement of atoms and photons by scattering photons off atoms in a BEC. Entanglement between atoms and photons has been considered by others and possible schemes have been proposed, \cite{Ruostekoski98}--$\!\!$\cite{Moore00}. In extending the existing work we have studied the factors which may limit the amount of entanglement achievable, in particular due to interactions between the atoms and between atoms and photons.
We consider weakly interacting atoms with just two internal states in a homogeneous BE-condensed state at zero temperature.
We consider in detail 2-photon Bragg transitions between different centre of mass motional states of the atoms in the internal ground state, with momenta $p$, $p'$ and energies $\hbar\ometil_{p}$, $\hbar\ometil_{p'}$ \cite{VectorNotation}. The intermediate state is taken to be an internal excited state. 
The initial state will be assumed to be a product of the weakly interacting condensate ground state and a coherently populated photon mode with momentum $k_1$. We then calculate the time evolution of the system. Under the assumption that the light induced coupling between the atomic modes is weak enough to neglect the depletion of the condensate due to this interaction. 
We envisage an experiment, where a cw laser with frequency $\ome_1$ and photon momentum $k_1$ is incident on a BE-condensate, where photons with frequency $\ome_2$ are scattered into modes $k_2$. In general, such an experiment will produce photons scattered in many different directions, together with atoms created in modes with the corresponding recoil momenta $\Del k=k_1-k_2$. However, we restrict our calculations to a single scattered photon mode, assuming that it may be singled out experimentally. In this way we define the atom-photon mode-pair, whose relative number squeezing we consider.  By adjusting the angle between $k_1$ and some fixed $k_2$, the frequency difference $\Del\ome=\ome_1-\ome_2$ can be tuned to the difference $\ometil_{p'}-\ometil_{p}$ of the atomic quasiparticle frequencies. This allows access of both the quasiparticle and the particle regimes of the centre of mass spectrum \cite{Ketterle01}.

Before we introduce the model in more detail we give a review of relative number squeezing and introduce a measure for it.
We would like to study the correlations between the atoms and the quantized photon mode. The two mode squeezing operator has the form
\be
\label{eq1}
  \oS(\zeta) = \exp\lk(-\zeta\oadg\obdg+\zeta^*\ob\oa\rk).
\ee
Here $\oa$ and $\ob$ are the bosonic creation operators in modes $A$ and $B$ respectively, and $\zeta$ is a c-number. 
The state resulting from this operator acting on the particle number vacuum state would be ideally squeezed, since particles are only generated in pairs. It may be written in the form
\be
\label{eq2}
  \oS(\zeta)\,|\,0_a;0_b\ra
  = \sech r \sum_{n=0}^\infty [-e^{i\phi}\tanh r]^n\,|\,n_a;n_b\ra,
\ee
where $\zeta=r\exp(i\phi)$, with $r$ and $\phi$ real and positive (cf.\ \cite{BarnettMiTQO}). This state is entangled and the distribution function over the different occupation numbers $n$ is thermal. 

The squeezing operator (\ref{eq1}) is, up to a diagonal phase matrix, equivalent to the time evolution operator describing the Bragg transitions we consider, if we restrict ourselves to the simplified case where one of the atomic centre of mass modes involved in the 2-photon Bragg transition is the condensate mode, $p=0$, and the other mode is $p=\Del k$. Atoms in mode $p=\Del k$ and photons in mode $k_2$ are thus produced in pairs, and on resonance, $\Del\ome=\ometil_{\Del k}$, 
the time evolution operator is essentially given by (\ref{eq1}).

The squeezing is measured in terms of the variance of the occupation number difference. For state (\ref{eq2}) this variance is zero,
\bea
\label{eq3}
   \lk[\Del(n_a-n_b)\rk]^2 
   &=& \la (\on_a-\on_b)^2\ra - \la\on_a-\on_b\ra^2
   \nonumber\\
   &=& (\Del n_a)^2+(\Del n_b)^2 
       -2\,\la\!\!\la\on_a\on_b\ra\!\!\ra
   = 0.
\eea
where $\on_a=\oadg\oa$, $\on_b=\obdg\ob$, $\la\!\!\la\on_a\on_b\ra\!\!\ra=\la\on_a\on_b\ra-\la\on_a\ra\la\on_b\ra$. 
We define a quantity which describes the relative number squeezing between modes $a$ and $b$:
\be
\label{eq4}
   \xi_3 := \lk[\Del(n_a-n_b)\rk]^2/\la\on_a+\on_b\ra.
\ee
This parameter is equivalent to the one defined in \cite{Wineland94} measuring the squeezing considered in \cite{Kitagawa93}. For independent classical, i.e.~coherently populated states, the correlation term $\la\!\!\la\on_a\on_b\ra\!\!\ra$ in (\ref{eq3}) is zero and the squeezing parameter is equal to $1$. The thermal state (\ref{eq2}) which has $\xi_3=0$ is maximally squeezed. Note that this state may also be considered as maximally entangled in the sense that the relative entropy $S_{rel}=\Tr_a[\rho_a\ln\rho_a]$ is at its maximum, where $\rho_a=\Tr_b[\rho_{a,b}]$ is the reduced density matrix obtained from the density matrix $\rho_{a,b}$ corresponding to the state (\ref{eq2}).
In the context of spin squeezing \cite{Wineland94} the parameter (\ref{eq4}) measures the variance of the 3-component $\oJ_3=\eh(\on_a-\on_b)$ compared to half the total value $J/2=\ev\la\on_a+\on_b\ra$ of an abstract angular momentum. It will be instructive to study also the squeezing of $\oJ_1=\eh(\oadg\ob+\obdg\oa)$ and $\oJ_2=\edzi(\oadg\ob-\obdg\oa)$ for which we define $\xi_i:=[\Del J_i]^2/(J/2)$, $i=1,2$.
 
There are two reasons why perfect squeezing $\xi_3=0$ is spoilt in an experiment: Collisional atomic interactions and photon (rescattering) processes different from those we above restricted our view upon.

To a first approximation, the ground state of an homogeneous interacting BE-condensed atomic gas, i.e.~the quasiparticle vacuum, may also be written as a state of the type given in (\ref{eq2}), with relative particle number squeezing between modes of opposite momenta $p$ and $-p$. Thus, even without photon scattering, the squeezing between the (empty) mode $k_2$ and atomic mode $p'=\Del k$ is spoilt, with a variance of the relative occupation being equal to the variance of the population of mode $p'=\Del k$, $[\Del (n^a_{k_2}-n^b_{\Del k})]^2=[\Del n^b_{\Del k}]^2\not=0$.

Moreover, one has to take into account that there are other processes of photon induced scattering into and out of the condensate mode. Specifically, an atom in mode $p=-\Del k$ may be transferred into the condensate together with a photon scattered from $k_1$ into $k_2$, and vice-versa. These processes appreciably affect the squeezing between $p'=\Del k$ and $k_2$ at longer times as our calculations presented below show. 

To study the time evolution of the squeezing we use different methods: We may either treat the effective two-photon coupling between the atomic c.m.~states as a perturbation to the Hamiltonian describing the free evolution of the interacting condensate and the photons and apply perturbative methods. Alternatively we may compute the time evolution in an exact way within a restricted Fock space. Due to the Bose-enhancement of scattering processes involving the condensate mode, we may restrict our view on the atom modes $\pm\Del k$. On resonance ($\Del\ome=\ometil_{\Del k}$) the Hamiltonian restricted to the subspace of these modes and the photon mode $k_2$ is periodic in time and we may use the Floquet-approach \cite{Floquet1883} to solve the Schr\"odinger equation.

We consider the homogeneous case, where the atomic motion is not confined by a trapping potential, but we assume that the system is confined in a box of volume $V$, and apply periodic boundary conditions. 
The homogeneous picture should be a good approximation for the central region of the large alkali condensates achieved in present day experiments. An extension of our methods to the case of trapped condensates is straightforward.

The effective Hamiltonian describing 2-photon Bragg transitions within the different centre of mass motional levels of the atom being in the internal ground state is obtained by adiabatic elimination of the intermediate state. It is written in terms of the operators $\oa$ and $\ob$, which satisfy standard bosonic commutation relations $[\oa_k,\oadg_{k'}]=\del_{kk'}$, etc., as $\oH = \oH_0 + \oHaf,$ where
\bea
\label{eq6}
  \oH_0 
  &=& \sum_{p,q\not=0}\,\lk[\cL_{pq}\obdg_{p}\ob_{q} +
      \eh\lk(\cM_{pq}\obdg_{p}\obdg_{q}+h.c.\rk)\rk] 
  \nonumber\\
  & &+\ \hbar\ome_2\oadg_{k_2}\oa_{k_2},\\
\label{eq7}
  \oHaf
  &=& \hbar\Ometil\sum_{p}\,
  \lk[
      \oadg_{k_2}\obdg_{p}\ob_{p-\Del k}e^{-i\ome_1t} 
      + h.c.\rk].
\eea
Here $\cL_{pq}=(\hbar^2p^2/2m+n_0U_0)\del_{pq}$ and $\cM_{pq}=n_0U_0\del_{p(-q)}$ are the coefficients of the quadratic Hamiltonian describing the motion of the colliding atoms of mass $m$. The collisional interaction has been approximated by the contact "potential" $V(\rv)=U_0\del(\rv)$, the coupling $U_0$ being related to the atomic scattering length $a$ by $U_0=4\pi a\hbar^2/m$.
Any dependence on the polarization of the photons is ignored.
We have neglected the AC-Stark-shifts, which would modify the diagonal terms in the basis of the particle operators and could easily be included when necessary in modelling an experiment.
The 2-photon Rabi frequency $\Ometil$ is given in terms of the dipole momentum $\dv_{ge}$ between the internal ground and excited states, of the positive frequency part of the electric field describing the laser mode 1, and the coupling $g_{ge}(k_2)$ of the photon mode 2 as  
\be
\label{eq7a}
  \Ometil = -2{|\dv_{ge}\cdot\cEv^\iKvar{+}(k_1)|\,g_{ge}(k_2)}/{\Del}.
\ee
$\Del=\ome_{k_1}-\ome_0$ is the detuning of the laser from the internal atomic frequency difference $\ome_0$. 
    
$\oH_0$, which includes interactions in the Bogoliubov approximation, may be diagonalized and written in terms of quasiparticle operators $\obet_p$ given by
\bea
\label{eq8}
  \obet_{p}
  &=& u_p \ob_p + v_p \obdg_{-p}.
\eea
The positive coefficients $u_p$, $v_p$ of the transformation are 
$u_p = (1-\alp_p^2)^{-1/2}$, 
$v_p^2=u_p^2-1$, with
$\alp_p = 1+y_p^2-y_p(2+y_p^2)^{1/2}$,
$y_p = |p|/p_0$,
$p_0 = \sqrt{8\pi an_0}$, $n_0$ being the condensate density.
For simplicity we neglect the change in $u_p$, $v_p$ by the depletion of the condensate due to photon scattering.

The Hamiltonian in terms of quasiparticle operators is
\bea
\label{eq11a}
  \oH 
  &=& \sum_{p}\,\hbar\ometil_{p}\obetdg_{p}\obet_{p} +
      \hbar\ome_2\oadg_{k_2}\oa_{k_2}
      \nonumber\\
\label{eq11b}
  & &\ +\ \hbar\Ometil\sum_{p}\,\lk[
      \oadg_{k_2}\obdg_{p}\ob_{p-\Del k}e^{-i\ome_1t}
      +\ h.c.\rk],
\eea
with $\ob_p=u_p\obet_{p}-v_p\obetdg_{-p}$ and
quasiparticle frequencies
\be
\label{eq12}
  \ometil_p 
  = (\hbar p_0^2/2m)\,y_p\sqrt{2+y_p^2}.
\ee
Since the recoiling atoms originate predominately from the condensate and have a momentum $p=\Del k$, we will focus on the squeezing between the photon mode $k_2$ and the atomic mode $\Del k$.
As inital state we choose $|\,t=0\ra=|\,\phi_0;0\ra$, defined by:
$  \obet_p|\,\phi_0;0\ra
= \oa_{k_2}|\,\phi_0;0\ra = 0$,
$  \ob_0|\,\phi_0;0\ra
= \sqrt{N_0}|\,\phi_0;0\ra$,
where $N_0$ is the number of atoms in the condensate.\\ 
  
In the following we present numerical results for the squeezing parameters $\xi_i$, $i=1,2,3$, for a realistic example of a sodium condensate.
We consider specifically a homogeneous condensate of $N=10^6$ $^{23}$Na atoms within a volume $V=10^{-8}\,$cm$^3$, which we choose such that the condensate density $n_0=10^{14}\,$cm$^{-3}$ is of the order which is typically observed in experiments. With $a=2.8\,$nm we thus have $p_0=2.65\cdot10^6\,$m$^{-1}$, $\hbar p_0^2/2m=(2\pi)\,1.5\,$kHz. 

\begin{figure}[tb]
\begin{center}
\epsfig{file={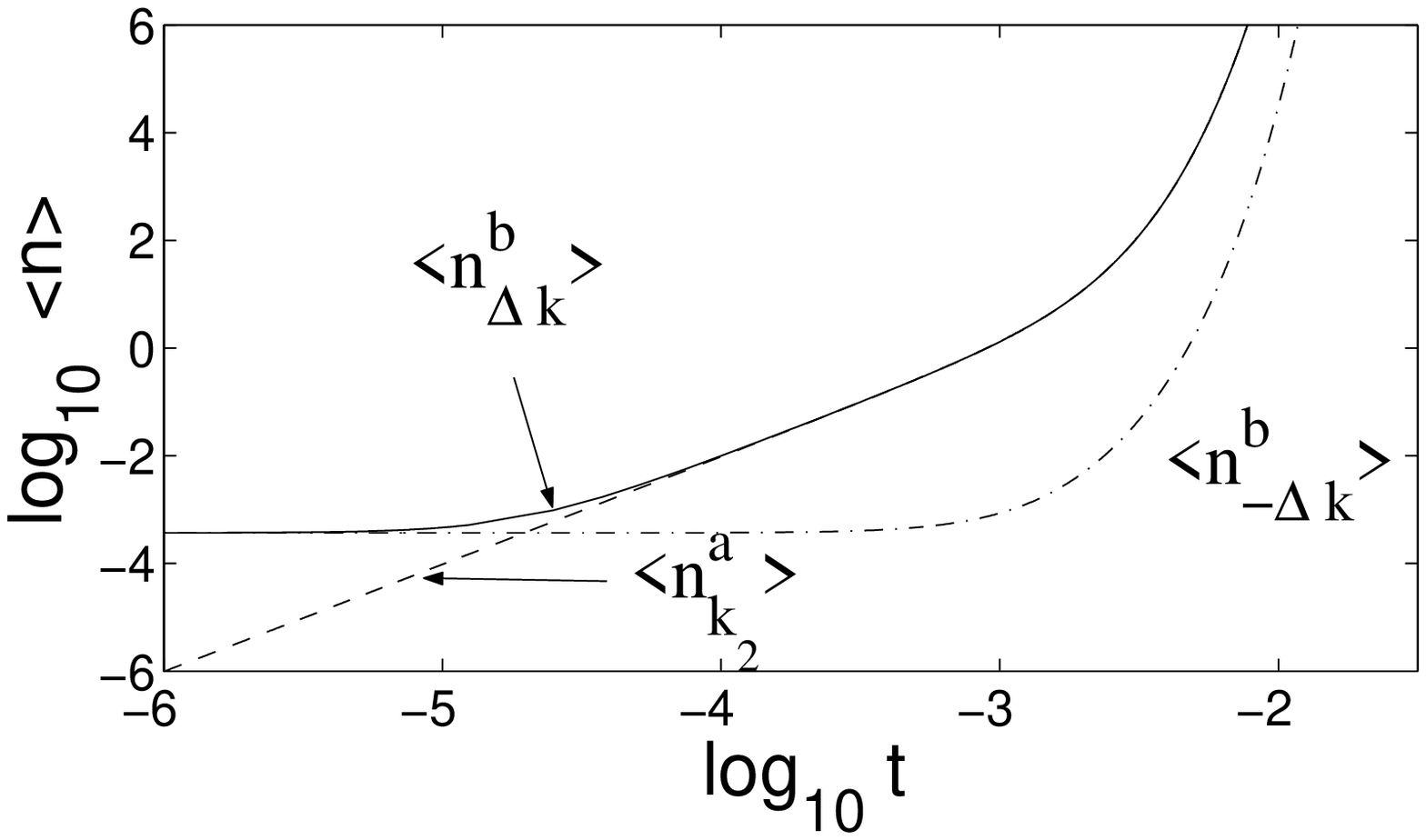},height=3.5cm,width=8cm,angle=0}
\caption{\label{fig2}Time dependence of the mean particle occupation numbers of the scattered photon mode, $\la\on^a_{k_2}\ra$, and of the modes of the recoiling atoms, $\la\on^b_{\Del k}\ra$, $\la\on^b_{-\Del k}\ra$.}.
\end{center}
\begin{center}
\epsfig{file={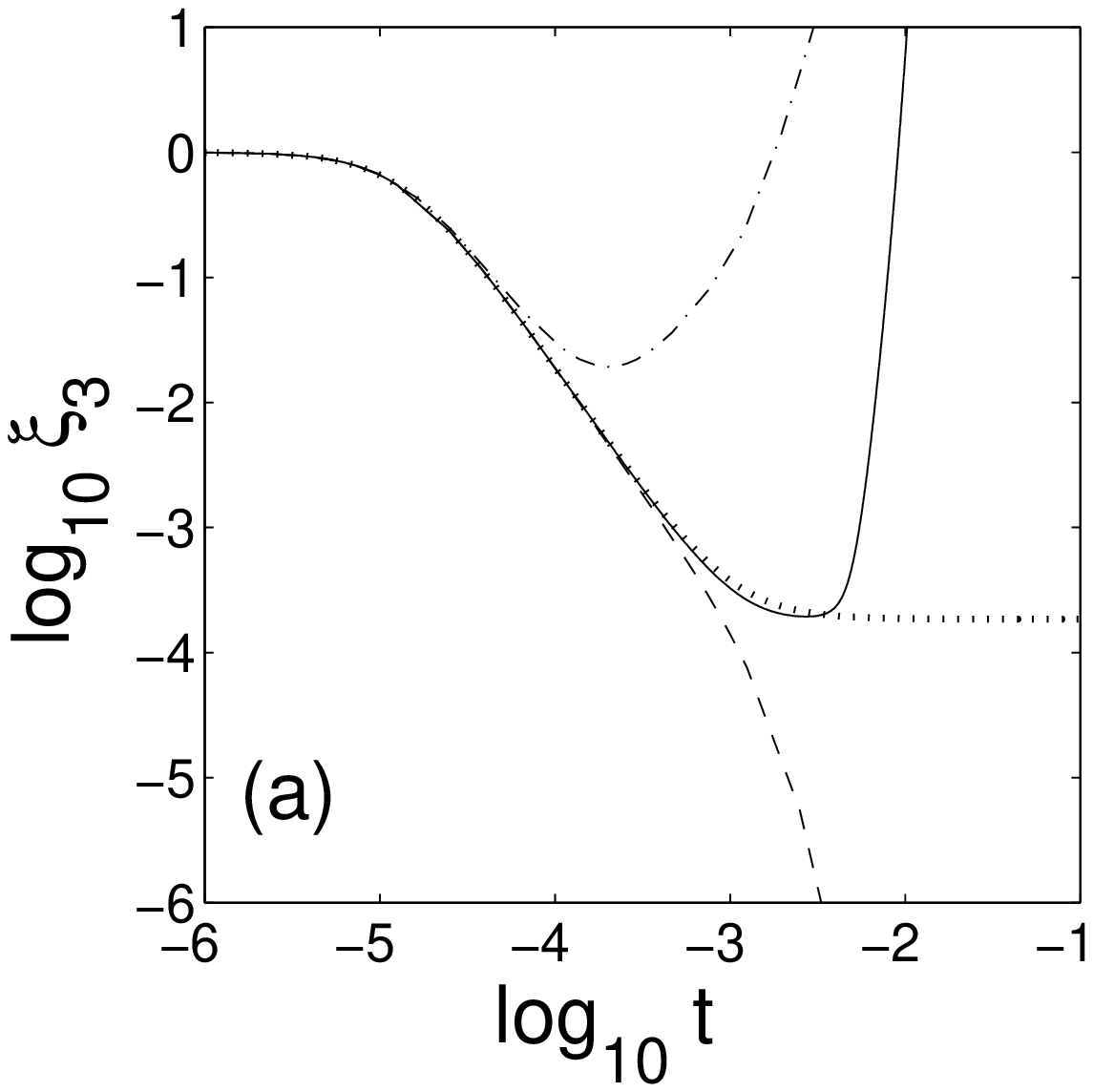},height=3.5cm,width=4cm,angle=0}
\epsfig{file={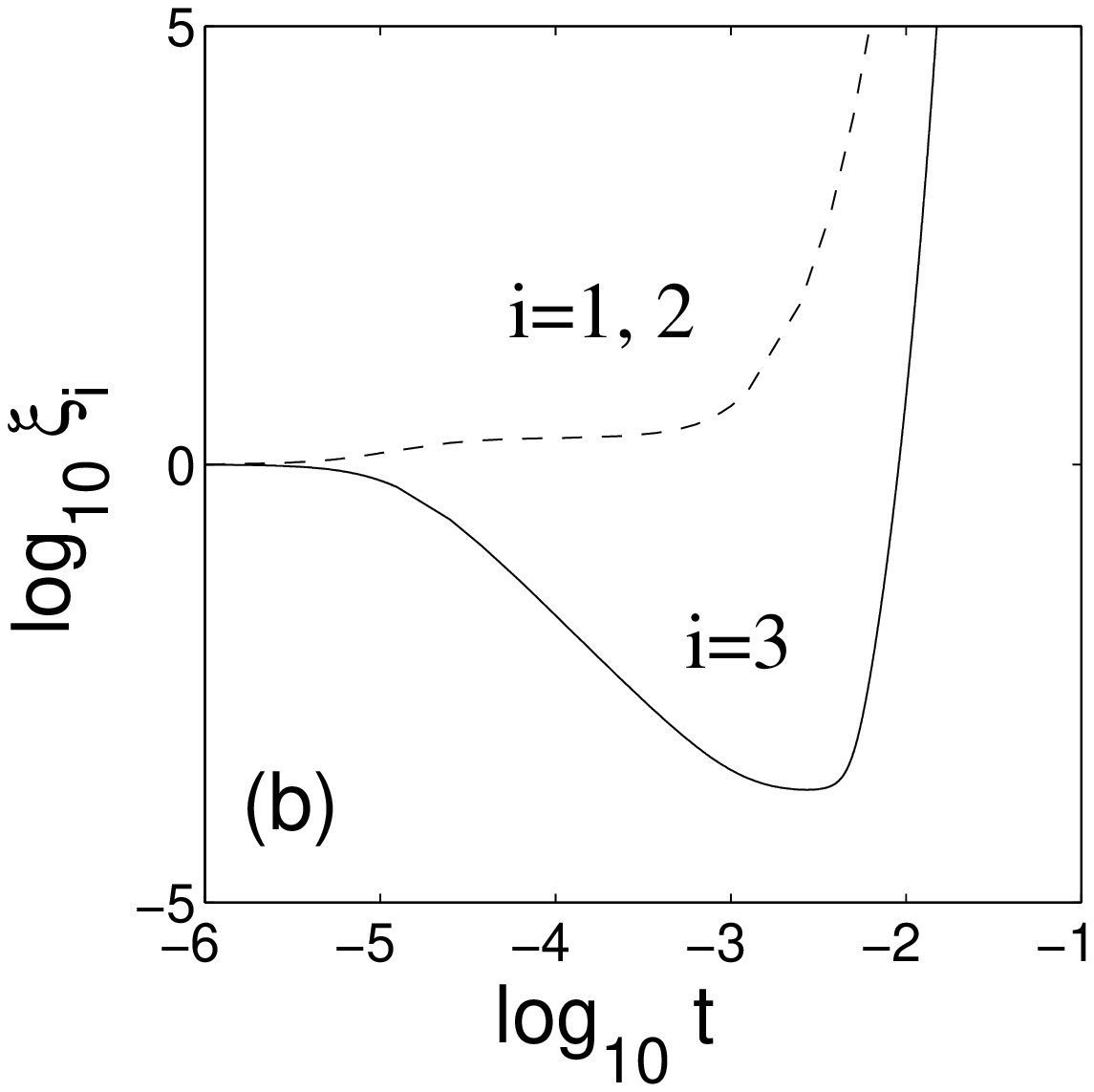},height=3.5cm,width=4cm,angle=0}
\caption{\label{fig3}(a) Time dependence of the squeezing parameter $\xi_3$: exact calculation (solid line), perturbative calculation (dotted line), energy  differences neglected (dash-dotted line), secondary recoiling mode $p=-\Del k$ neglected (dashed line). (b) Time dependence of the squeezing parameters $\xi_i$.}
\end{center}
\end{figure}

We consider the case where the photon recoil energy $\Del\ome=\ome_1-\ome_2$ is tuned to the resonance $\Del\ome\simeq\ometil_{\Del k}$. In Fig.~\ref{fig2} we plot the logarithm of the occupation numbers $\la\on^a_{k_2}\ra$, $\la\on^b_{\Del k}\ra$, $\la\on^b_{-\Del k}\ra$, for $y_{\Del k}=5$, i.e.~for coupling into the particle regime of the spectrum (\ref{eq12}), where $\ometil_p\simeq \hbar p^2/2m$. The curves have been computed using the Floquet approach, by restricting the calculation to the subspace of these three modes. This is possible since on resonance, the interaction Hamiltonian is periodic in time, with period $T=\pi/\ometil_{\Del k}$.
We have chosen a two photon Rabi-frequency of $\Ometil=1\,$s$^{-1}$. Initially the atomic occupation is $\la\on^b_{\Del k}(t=0)\ra=v_{\Del k}^2$ due to collisional interactions. Only for $t\gsim0.5\,$ms the photon and atom number occupations  are of the same magnitude, $\la\on^a_{k_2}\ra\simeq\la\on^b_{\Del k}\ra$. We have checked that for times shorter than $\sim1\,$ms the curves are equivalent to those obtained from a perturbative calculation to order $\Ometil^2$ within the whole Fock space of atomic modes. For $t\gsim1\,$ms a non-perturbative rise becomes significant. For $\la\on^b_{\pm\Del k}\ra$ approaching $N_0=10^6$ however, the calculation reaches its limit of physical applicability since we have neglected the depletion of $N_0$. The range of times where the calculation is valid may of course be increased by choosing a smaller $\Ometil$.

Fig.~\ref{fig3}a shows the squeezing parameter $\xi_3$. For times $t\lsim10\,\mu$s it is close to $1$, i.e.~there is essentially no squeezing compared to the classical case of independent modes. This is due to the fact that the relative atom-photon number variance is still dominated by the initial thermal variance of the atom mode population, $[\Del n^b_{\Del k}(t=0)]^2=\la\on^b_{\Del k}\ra(\la\on^b_{\Del k}\ra+1)=v_{\Del k}^2(v_{\Del k}^2+1)$, which, in the particle regime where $v_{\Del k}^2\ll1$, resembles the variance of a coherent state. (For recoil momenta in the quasiparticle regime, $\Del k<p_0$ the initial value of $\xi_3$ would be considerably larger than 1.) For $t\gsim10\,\mu$s the squeezing starts to improve due to the atom-photon pair production, but only until $t\simeq1\,$ms. The perturbative calculation to 2nd order in $\Ometil$ gives a settling of $\xi_3$ to $\simeq v_{\Del k}^{2}\simeq10^{-4}$ (dotted line). The exact Floquet calculation within the restricted Fock space results in $\xi_3$ rising again for $t\gsim3\,$ms (solid line). It is interesting to observe what causes the squeezing to be limited at large times. Neglecting the 2-photon Bragg transitions between $p=-\Del k$ and $p'=0$ results in good squeezing where the time evolution is described by (\ref{eq1}) at long times (dashed line). Including the processes involving the mode $p=-\Del k$ but neglecting the energies $\Del\ome$ and $\ometil_{\pm\Del k}$ means that the time evolution is described by the operator
\be
\label{eq13}
  \oS(\zeta,\vartheta) 
  = \exp\lk(-\zeta\,\oadg\obdg_{\Del k}
            -\vartheta\,\oadg\ob_{-\Del k}+h.c.\rk),
\ee
where $\zeta=\vartheta=i\Ometil\sqrt{N_0}t$. In this case the squeezing between $p'=\Del k$ and $k_2$ is reduced already for $t\gsim0.1\,$ms (dash-dotted curve). It is therefore the energy separation of the motional levels from the condensate which lowers the degree to which the squeezing is suppressed and yields the time evolution as shown by the solid line.

The three mode operator (\ref{eq13}) also shows why the two-mode squeezing between $p=\Del k$ and $k_2$ decreases at large $t$. A change of the occupation number difference $n^a_{k_2}-n^b_{\Del k}$ by minus one is equivalent to the creation of a particle in mode $p=-\Del k$. Thus the variance $[\Del (n^a_{k_2}-n^b_{\Del k})]^2$ must be the same as the variance $(\Del n^b_{-\Del k})^2$ of the population of the $p=-\Del k$ mode and numerical calculations reflect this. Hence, the creation of a perfectly squeezed state w.r.t.~modes $p'=\Del k$, $k_2$ requires the mode $p=-\Del k$ to be in a number state. This is physically realistic only when the mode is empty. However the atomic collisions cause any mode with $p\not=0$ to be occupied to a certain extent, with a thermal variation $(\Del n^b_{p})^2=\la\on^b_{p}\ra(\la\on^b_{p}\ra+1)$, and this limits the squeezing right from the beginning. The photon induced transitions then lead to an increase of $\la\on^b_{-\Del k}\ra$ with time, with $(\Del n^b_{-\Del k})^2$ remaining a thermal variance. For the collisionless case ($a=0$) a similar squeezing deterioration at large times has been found in \cite{Moore99}.

We still note that the squeezing of the difference of the {\it quasi}particle
occupation number of the atomic mode $\Del k$ and the particle occupation of the photon mode $k_2$ is at maximum at time zero. But due to the population of the quasiparticle mode $-\Del k$, the time evolution of the quasiparticle squeezing is similiar to that of the particle squeezing for times $t\gsim3\,$ms.

In the context of spin squeezing, we found analytically that within the restricted Fock space of the modes $\pm\Del k$, $k_2$, we have $\la\oJ_1\ra=\la\oJ_2\ra\equiv0$ for all times. Thus the mean spin vector $\la\oJv\ra$ points in 3-direction. Fig.~\ref{fig3}b then shows, that although the squeezing of all three spin components becomes worse for large times compared to the case of coherent modes, there is a strong relative squeezing of several orders of magnitude of the 3-component compared to the 1- and 2-components which are of equal size.  

It might be argued that recollision processes of photons created in mode $k_2$ may be avoided by choosing an optically thin BE-condensate, which the photons are leaving before they may recollide. Such a scheme has been proposed in \cite{Moore00}. This can help to avoid the squeezing being spoilt at large times. We estimate the fraction of atoms and photons in the modes $\Del k$ and $k_2$ being rescattered before leaving the condensate as $r_x=\sigma_x n_0 l$, where $\sigma_x$ is the total cross section for atoms, $\sigma_{\mathrm{at}}=8\pi a^2$, or photons, $\sigma_{\mathrm{ph}}=(3/2\pi)(\lam\gamma/\Del)^2$, and $l$ the width of the sample. With $\gamma=0.6\cdot10^8\,$s$^{-1}$ being the natural linewidth of the sodium $\lam=589\,$nm line, and a typical detuning of $\Del=1\,$GHz we obtain for our example condensate $r_{\mathrm{at}}=42\%$ and $r_{\mathrm{ph}}=3\%$. 
This shows that although the photon rescattering plays a relatively small role, there will be further limitations of squeezing arising from loss in the scattered atomic mode, which we have not included into our calculations so far.
Despite this our results show that there is a substantial initial limitation of squeezing due to the collisional interactions between the atoms, which are naturally present in the gas before the laser is switched on.


%% file: Letter.bbl
\begin{references}
\bibitem{VectorNotation}
In this letter, vectors $\pv$ are denoted by $p\equiv(p_x,p_y,p_z)$.
\end{references}
